# Cooper pairs or Individual electrons? Based on the magnetic flux quantization and the Tc scaling law of superconducting thin films

Xiang Wu

**Abstract:** We look back on the turning point of the concept of individual electron carrier and paired electrons carrier in superconductivity history, look back on the magnetic flux quantization interpretation and experiment, reference the new discovery of new phenomena including the hc/e magnetic flux periodicity and the scaling law of transition temperature of superconducting thin films. The conclusion is: In a circuit including only one electron, the magnetic flux period is hc/e in general, but hc/(2e) in the first step starting with zero flux. Early experiments had not observed hc/e periodicity because multiple electrons do each first step starting with zero flux. According to the new experimental explanation, the theory of individual electron carrier is proposed. It integrates with the energy band theory and the Hubbard model. In the frame of energy band theory and electron correlation, we do not add Cooper pair for the carrier, we can use the energy band theory to explain the phenomenon of superconductivity. It needs the Hubbard model when a superconductor is a Mott insulator. The theory makes the whole image of super-insulators, insulators, semiconductors, conductors, superconductors transition smoothly and continuously. The transition to fractional quantum Holzer effect is continuous and smooth.

**Keywords:** superconductivity; magnetic flux; fractional quantum Holzer effect; energy band theory; BCS theory

# 0 Background

In 1961, Deaver and Fairbank's experiments [1], Doll and Näbauer's experiments [2], the experimental data and the theoretical speculation have been treated as formal explanations to now. The idea is that the magnetic flux of a superconducting circuit increases by hc/(2e). According to the experimental interpretation at that time, the idea is that a superconducting carrier is paired electrons. It supports the BCS theory.

The first discovery of high transition temperature superconductivity in 1986 exceeded the prediction of early BCS theory. Since then, some difficulties have been solved by modifying the mechanism of electron pairing. Up to now, BCS theory has not fully explained all high Tc superconductivity phenomena [3].

The latest experimental results of I. Boovivich, X. He, J. Wu, A. T. Bollinger in 2016 opposed to the prediction of BCS theory [4]. The relationship between carrier number and transition temperature is contrary to BCS theory prediction.

In 2016, TaoYong's superconducting thin films temperature scaling law[5] was just in line with the experimental results of the paper[4].

The other background in the field of superconductivity includes below.

Before 1953, Cheng Kai-Jia and Born proposed the double band theory of superconductivity [6]. Professor Cheng did not agree with the BCS theory.

In 2007, Heersche, Jarillo-Herrero, Oostinga, Vandersypen & Morpurgo used the Andreev reflection concept to explain the transition of double-electron and single-electron in superconducting-graphene-superconducting junctions [7].

In 2008, Tzu-Chieh Wei, Paul M. Goldbart measured the hc/e magnetic flux period in the superconducting ring experiment and summarized the experience that the smaller ring, the easier it is to show hc/e magnetic flux period [8].

In 2008, Loder, Kampf, Kopp, Mannhart, Schneider, Barash cited a variety of magnetic flux period as hc/e, the magnetic flux change period is not always hc/(2e) [9].

In 2008, Vinokur, Baturina, Fistul, Mironov, Baklanov & Strunk discovered the phenomenon of super-insulation. It is a phenomenon with [10] superconductivity, which is completely opposite to superconductivity. The resistance of a superconductor is infinitely close to zero, and the conductance of a super-insulator is infinitely close to zero. Using mathematical concepts to say that the resistance of a superinsulator is meaningless, because it is a zero divisor.

In 2009, Hirsch issued many doubts on BCS theory [11].

In 2012, Cao Tian-De published his views on the Josephson effect. The Josephson effect is not the tunneling of electron pairs, but also the single electron passing through [12]. Cao pointed out the disassembly, tunneling, and the pairing of electrons. The phenomenological theory regards "dismantling-tunneling-pairing" as the equivalent "tunneling of electron pairs", which is a coincidental success but conceals the physical essence.

In 2013, Rosenberg and Agassi gave the loop electron flux quantization [13] like Mobius rings. The predicted flux period is hc/(4e).

In April 2016, Hamidian, Edkins, Sang Hyun Joo, Kostin, Eisaki, Uchida, Lawler, Kim, Mackenzie, Fujita, Jinho Lee, Séamus Davis first experimentally measured electron (pair) density waves in superconductors [14].

In May 2016, Van Zanten, Basko, Khaymovich, Pekola, Courtois, Winkelmann experimentally demonstrated the superconducting single electron tunneling effect [15]. Unlike the single electron tunneling effect with resistance. It is generally believed that the Josephson effect with only two electrons is actually a single electron.

**1 Status in the study**

Before this paper, we launched a discussion in the public. Based on at least two rare and affirmative experimental results since 2008, the appearance of superconducting carriers has been deduced. The carrier is not the Cooper pair that was originally described in 1957 BCS theory. We propose to modify one description on the basis of BCS theory: Cooper pair is not the carrier. Responsible for the carrier is still not paired electrons. Carriers are free/individual electrons.

The energy band theory including free electron approximation and the interaction between electrons can be used to understand superconductivity by band diagrams and electron filling rules. The modified description is equivalent to the old description. In terms of conventional superconductivity, when a stable Cooper pair exists, the Cooper pair will not be dismantled and exchanged with the carrier electrons, that is, the material exhibits superconductivity. For high Tc superconductor and conventional superconductor, the material exhibits superconductivity when the fully filled band electrons is not dismantled and exchanged with the current-carrying electrons, and/or the valence band electrons are not exchanged with the current-carrying electrons.

By now, expanded discussions have taken shape. This paper focuses on explaining the questions that must be answered behind this understanding. In order to eliminate doubts and even disdain emotions.

## 2 Common questions and answers

By insisting that Cooper pairs are carriers, scholars may ask:

How can we explain why the flux quantization experiment has shown that superconducting carriers are advancing in pairs?

We answer:

Back in 1961, when the quantization of magnetic flux was made and the artificial conclusion was drawn, it would be fair to assume that another formulation was proposed before the academic community reversed to the point that Cooper pairs were carriers, and Cooper pairs carriers would be worth criticizing. BCS theory uses the concept of phonon, proposed Cooper pair, clearly explained the gap, but said that magnetic flux quantization experiment "proves" the superconducting carriers in pairs, be contrary to another basic principle of quantum mechanics - Fermion obeys Dirac statistics.

In experiments, people pay attention to the motion shown by electrons, which is described by Bloch electrons. The global motion of Bloch electrons also obeys the basic principle of quantum mechanics. The appearance of Bloch electrons is also fermions. In a closed loop (or solenoid), the wave function of the Bloch electron constructs an odd number of antinodes to conform to Dirac statistics. This shows that the phase of the wave function of the electron changes the odd number of $\pi$ every loop.

The 1961 literature is based on the even number of $\pi$. Relevant conclusion has been widely adopted so far. One sentence seems to be very perfunctory today. "Because the wave function is a single value function, the electron phase changes the integral number of $2\pi$ around a circle." People who study quantum mechanics have never said that Dirac's statistical function is not a single-valued function. In other words, under the premise of understanding the properties of fermions, People who study quantum mechanics didn't say that the function of returning around two circles is not a single-valued function.

By insisting that Cooper pairs are carriers, scholars may ask:

How to explain the exact results of the magnetic flux period hc/(2e)?

We answer:

The hc/e magnetic flux period has been verified by new experiments in paper [8]. We explain the cause of hc/e and give experimental predictions. For each carrier in the loop, the first step to contribute magnetic flux from zero is hc/(2e), and the next step is hc/e. Early experiments did not observe the hc/e flux period because there was a considerable number of carriers in the loop. It is easy to observe the hc/e flux when the carrier number is reduced. The experiment [8] cited should be regarded as the reason.

By insisting that Cooper pairs are carriers, scholars may ask:

Cooper pairs of supercurrent is adopted precisely because the unpaired free electrons are scattered by the lattice vibration to produce resistance. If we talk about free electron current carrying, then scattering cannot be evaded. What is the explanation?

We answer:

Is scattering the cause of resistance? It's everybody's question. Where is the conclusive evidence? The lack of empirical evidence is only depicted in the brain. Today, we realize that it is impossible to simulate a large number of carriers scattered by phonons (lattice vibrations) and transfer energy to phonons. Rather than simulating a single locality, the correct simulation is a large number of carriers and phonons, long enough to evolve.

To be exact, we point out that the exchange energy of carriers and phonons is in dynamic equilibrium, that is, the heat balance. If we regard superconductor as carriers group and ions group. The temperature of the carriers group is equal to the temperature of the ions group.

In addition, the performance of Bloch electrons should also be subject to translational symmetry. After changing the inertial coordinate system, the laws of physics remain unchanged. In the inertial coordinate system following Bloch global walking, Bloch electrons are not moving.

By insisting that Cooper pairs are carriers, scholars may ask:

Since scattering is classified as superconductivity, how to explain resistivity?

We answer:

That's the problem. Resistivity and superconductivity are a pair of complementary concepts. If the explanation of resistivity is clear, then the explanation of superconductivity has not yet been pending. We point out that the cause of resistivity which is generally accepted is actually two cases, but not one.

In two cases, one case is that the carrier electron is still the carrier after the collision between the carrier electron and the ion, the other case is that the carrier electron is not a carrier after the combination of the carrier electron and the ion. So far, most scholars believe that resistance occurs in both cases. We pointed out that the case in which the carrier electrons are still carriers after the collision with the ions is actually superconducting and the resistance is zero. In view of the fact that the carrier electrons are not carriers after binding with the ion, the action of the electron being bound by the ion has to give energy, and this part of energy cannot be retrieved as it is when the electron becomes the next carrier.

After excluding the cause of a resistance and attributing it to the superconductivity explanation, the two concepts of resistivity and superconductivity are clearly separated and complementary to each other.

## 3 Experimental analysis

Take the [1][2] experiments in 1961 as an example to reanalyze. The experiment can be described as follows: when the external magnetic field provides magnetic flux through the superconducting closed loop, the induced magnetic flux of the superconducting ring changes from 0 to $hc/(2e)$ when the external magnetic flux increases from 0 to above $hc/(4e)$; when the external magnetic flux decreases from above $3hc/(4e)$ to lower than $3hc/(4e)$, the induced magnetic flux of the superconducting ring changes from $hc/e$ to $hc/(2e)$ . Each jump occurs at a position that provides a magnetic flux equal to half an integral number of flux quantum, that is, half an integral multiple of $hc/(2e)$. At any time, the induction flux is equal to the integer number of flux quantum, that is, the integral number of $hc/(2e)$.

According to the basic principles of quantum mechanics, electrons are fermions, when an electron appears in the same area twice, the phase must be odd times of $\pi$ (using radian), even for electrons in superconducting loops. We make the derivation process of superconducting magnetic flux quantization.

Books usually start with formula (1). $j_S$ is the current density provided by superconducting electrons. It should be pointed out that according to the description of existing books, a large number of superconducting electrons occupy a single quantum state together, so the matter wave of each electron is the same, so formula (1) can be obtained. However, inheriting the band theory requires that each electron occupy a different quantum state. We point out that the matter wave examined here is not a matter wave in the quantum state of the electron, but a matter wave represented by the global motion of the Bloch electron. Due to the limitations of standing wave conditions, a large number of Bloch electrons obey the uniform wave function, and thus formula (1) can be obtained.

$$j_S = -\frac{hq}{2\pi m_S} n_S(r) \nabla \theta(r) - \frac{q^2 n_S(r)}{m_S} A \qquad (1)$$

The part of the superconducting ring that leaves the depth of penetration on the surface $j_S = 0$. Formula (1) is integrated by a loop, and equation (2) is obtained.

$$\oint_C A \cdot dl = -\frac{h}{2\pi q} \oint_C \nabla \theta \cdot dl \qquad (2)$$

The early theory held that the wave function was a single-valued function and that the phase difference could only be an even number of $\pi$ after the electron went around the loop. But now, we're pointing out that the phase difference can only be an odd number of $\pi$. There's formula (3).

$$\oint_C \nabla \theta \cdot dl = n\pi \quad (n = 0,1,3,5,7...) \qquad (3)$$

According to Stokes theorem, the line integral of equation (2) is transformed into a surface integral, and there is equation (4).

$$\oint_C A \cdot dl = -\oint_S \nabla \times A \cdot dS = \Phi \qquad (4)$$

$S$ is the area around the superconducting ring, and $\Phi$ is the flux that's passing through $S$. From formula (2), we get (5).

$$|\Phi| = n\frac{h}{2q} = n\phi_0 \quad (n = 0,1,3,5,7...) \qquad (5)$$

Among them

$$\phi_0 = \frac{h}{2q} \qquad (6)$$

The magnetic flux through the superconducting ring is quantized, odd times of $\phi_0$.

Equation (6) is different from the conclusion of existing books $\phi_0 = \frac{h}{q}$.

In the existing books, if q=2e, $\frac{h}{q}$ is in line with the experimental value $2.07 \times 10^{-15}$ Wb.

According to equation (6), if q=e, $\phi_0$ is in line with the experimental value $2.07 \times 10^{-15}$ Wb.

It can be seen that when the phase difference is equal to $\pi$, the quantum unit of magnetic flux is equal to hc/(2e), but q = e, which conforms to the single electron explanation. When the phase difference is greater than $\pi$, the magnetic flux jump period is equal to hc/e.

We point out that there is more than one electron in the original experiment. It is predicted that if there are only one or a few electrons, the hc/e hopping period can be observed after all electrons have realized the first step. [8][9] pointed out that the magnetic flux jump period is hc/e, which coincided with this view.

# 4 Theoretical application

The theory of individual electrons superconductivity also agrees with the Cooper pair and its energy gap. Cooper pair and its energy gap are confirmed by experimental phenomena. Including isotopic effect, it affirms the correctness of Cooper pairs and phonon concepts. The difference between the individual electrons superconductivity theory and the BCS theory is that a Cooper pair is two fully filled band electrons and the Cooper pair is not conductive. The theory of individual electrons superconductivity emphasizes that superconducting due to the prohibition of Cooper pairs and single-electrons converting to each other. For materials that are difficult to point out Cooper pairs, the individual electrons superconductivity theory emphasizes that superconducting due to the prohibition of fully filled band electrons and half filled band electrons converting to each other, and/or the prohibition of valence-band electrons and conduction electrons converting to each other. All experimental observations are under the same interpretation.

It is obvious that the super-insulator and superconductor are explained by individual electron carriers. Below the critical temperature, if there are free electrons (half filled band electrons) in the conduction band, and there is no energy exchange across the energy levels, then the material is a superconductor. Under the critical temperature, If there are no free electrons in the delocalized band, and the electrons in the valence band can not enter the conduction band (or the fully filled band electrons cannot be converted into half filled band electrons), then the material is a superinsulator. Correspondingly, ordinary insulators are always somewhat conductive, indicating that there are a few instantaneous free electrons in the conduction band.

As far as the current understanding of high Tc superconductors and topological insulators is concerned, superconducting materials have both super-insulating and superconducting phases. Especially the superconducting phase has attracted much attention. It becomes a macroscopic superconducting material when the superconducting phase changes from dispersed independent to contiguous.

The individual electrons superconducting theory makes the whole image of super-insulators, insulators, semiconductors, conductors, superconductors transition smoothly and continuously.

Super-insulators: There are no electrons at the delocalization level, or there are electrons at the delocalization level but they are a fully filled band, and there is no transient damage to the above characteristics.

Insulator: Delocalization energy levels generally have no electrons, or delocalization energy levels have electrons but are generally fully filled band electrons, locally there is a transient destruction of the above characteristics, but can not form a continuous conductor.

Semiconductor: The transition state as an insulator and conductor.

Conductor: There are a large number of free electrons in the delocalized energy level, and free electrons are not fully filled band electrons, so they conduct electricity. The resistance effect is caused by the exchange of free electrons with valence electrons or fully filled band electrons.

Superconductor: There are free electrons in the delocalized energy level, and free electrons are not fully filled band electrons, so they conduct electricity. The zero resistance effect is caused by no free electrons being exchanged with valence electrons or fully filled band electrons.

The above five material states can be converted to one another at different temperatures. Superconductors turn into insulators, semiconductors, and conductors in turn when the temperature rises and superconductors turn into conductors when the temperature rises.

The above five material states can be converted into one another under different pressures. The smaller the pressure, the more inclined the insulator or superinsulator, the greater the pressure, the more inclined the conductor or superconductor. The reason is that, as the distance between atoms is forced to shrink, the position of the Fermi plane gradually shifts to the position of the electron at the highest level. When there are free electrons in the conduction band, it becomes a conductor or a superconductor, and when there are no free electrons in the conduction band, it becomes an insulator or a super-insulator.

The state of the above five materials is dependent on the band properties of the material at the same temperature and pressure. People expect superconductivity at room temperature under atmospheric pressure.

The individual electrons theory can also explain fractional quantum Holzer effect. Fractional quantum Hall effect does not need to be realized in superconductors. It only needs very pure materials, even if the materials are resistive. Therefore, the phenomena of a conductor, the phenomena of superconductor and the picture of a smooth continuous transition reflect each other.

The fractional quantum Holzer effect is quantum Holzer resistance divided by the quantization of fractions. The denominator of a fraction is odd, and the numerator is an integer. Because the phase of electrons in the loop is odd times of $\pi$, electrons can only form odd antinodes in a homogeneous medium. Electrons can only appear on paths that allow odd numbered antinodes. The experiment of fractional quantum Hall effect usually adopts constant current and calculates Hall resistance by measuring voltage. Therefore, under constant current conditions, there are two voltages per two antinodes. It shows that there are two Holzer resistance in each of the two antinodes. This means that an odd number of the voltage leads to the quantization of the "quantum Hall resistance" divided by the fraction, where a denominator is an odd number. In addition, numerator are integers and are easy to understand. More than one electron involves more than one current, so the numerator is an integer.

## 5 Theoretical prediction

Using STM to verify the electron density wave in the superconducting closed loop, we will get the conclusion that the number of odd antinodes. Because it is an odd antinodes, which is consistent with the theory of individual electrons superconductivity. It should not be described as a Cooper pair density wave.

The magnetic flux quantization experiment of superconducting will show a step of hc/e period after all carriers begin to carry current. The smaller the number of carriers is, the easier it is to find hc/e. The experiment of bismuth single crystal superconductivity in 2008 is recommended.